\documentclass[%
 reprint,
 amsmath,amssymb,
aps,
]{revtex4-1}

\usepackage{graphicx}
\usepackage{dcolumn}
\usepackage{bm}
\usepackage{braket}

\usepackage{amsmath}
\usepackage{amsthm}
\usepackage{amssymb}
\usepackage{braket}
\usepackage{graphicx}
\usepackage{bm}
\usepackage{url}
\usepackage{hyperref}
\usepackage{dcolumn}
\usepackage{complexity}
\usepackage{lmodern}
\usepackage{natbib}
\usepackage{color}
\usepackage{booktabs}
\usepackage{epstopdf}
\usepackage{siunitx}
\usepackage{subfigure}
\usepackage{comment}
\usepackage{cancel}
\usepackage{resizegather}
\usepackage{xcolor}

\begin{document}

\preprint{APS/123-QED}

\title{Offset Simultaneous Conjugate Atom Interferometers}

\author{Weicheng Zhong}
 \email{zhongw@berkeley.edu}
\author{Richard H. Parker}
\author{Zachary Pagel}
\author{Chenghui Yu}
 \altaffiliation[Present address: ]{Department of Physics, University of California San Diego, 9500 Gilman Drive, La Jolla, CA 92093, USA}
\author{Holger M\"uller}
 \altaffiliation[Also at: ]{Molecular Biophysics and Integrated Bioimaging, LawrenceBerkeley National Laboratory, One Cyclotron Road, Berkeley,CA 94720, USA}
\affiliation{Department of Physics, University of California, Berkeley, California 94720, USA}

%

\date{\today}

\begin{abstract}

Correlating the signals from simultaneous atom interferometers has enabled some of the most precise determinations of fundamental constants. Here, we show that multiple interferometers with strategically chosen initial conditions (``offset simultaneous conjugate interferometers" or OSCIs) can provide multi-channel readouts that amplify or suppress specific effects. This allows us to measure the photon recoil, and thus the fine structure constant, while being insensitive to gravity gradients, general acceleration gradients, and unwanted diffraction phases - these effects can be simultaneously monitored in other channels. 
An expected 4-fold reduction of sensitivity to spatial variations of gravity (due to higher-order gradients) and a 6-fold suppression of diffraction phases paves the way to measurements of the fine structure constant below the 0.1-ppb level, or to simultaneous sensing of gravity, the gravity gradient, and rotations. 


\end{abstract}

\pacs{Valid PACS appear here}

\maketitle


Atom interferometers have been used for inertial sensing \cite{InertialChu, InertialBouyer, InertialKasevich, InertialLandragin, InertialShauYu}, testing Einstein's equivalence principle \cite{EEPWeitz, EEPRasel, EEPTino, EEPZhan, EEPHu}, measuring Newton's gravitational constant $G$ \cite{GKasevich, GTino} and searching for dark sector particles \cite{ChameleonBE, DarkEnergyPH}. Running two interferometers simultaneously and interrogating them with the same laser (Simultaneous Conjugate atom Interferometers, or SCIs) has proven to be a successful approach to cancel several systematic effects \cite{GammaTino, AlphaRP}. This cancellation by two SCIs is important, but is not perfect. Many common-mode effects, such as vibrations and uniform gravitational acceleration, will cancel out by extracting the differential phase between these two interferometers. However, gravity gradients and diffraction phases \cite{BraggKasevich, BraggChu}, which can be thousands of times larger than the needed final accuracy, remain.



\begin{figure}[ht]
  \centering
  \includegraphics[width=0.458\textwidth]{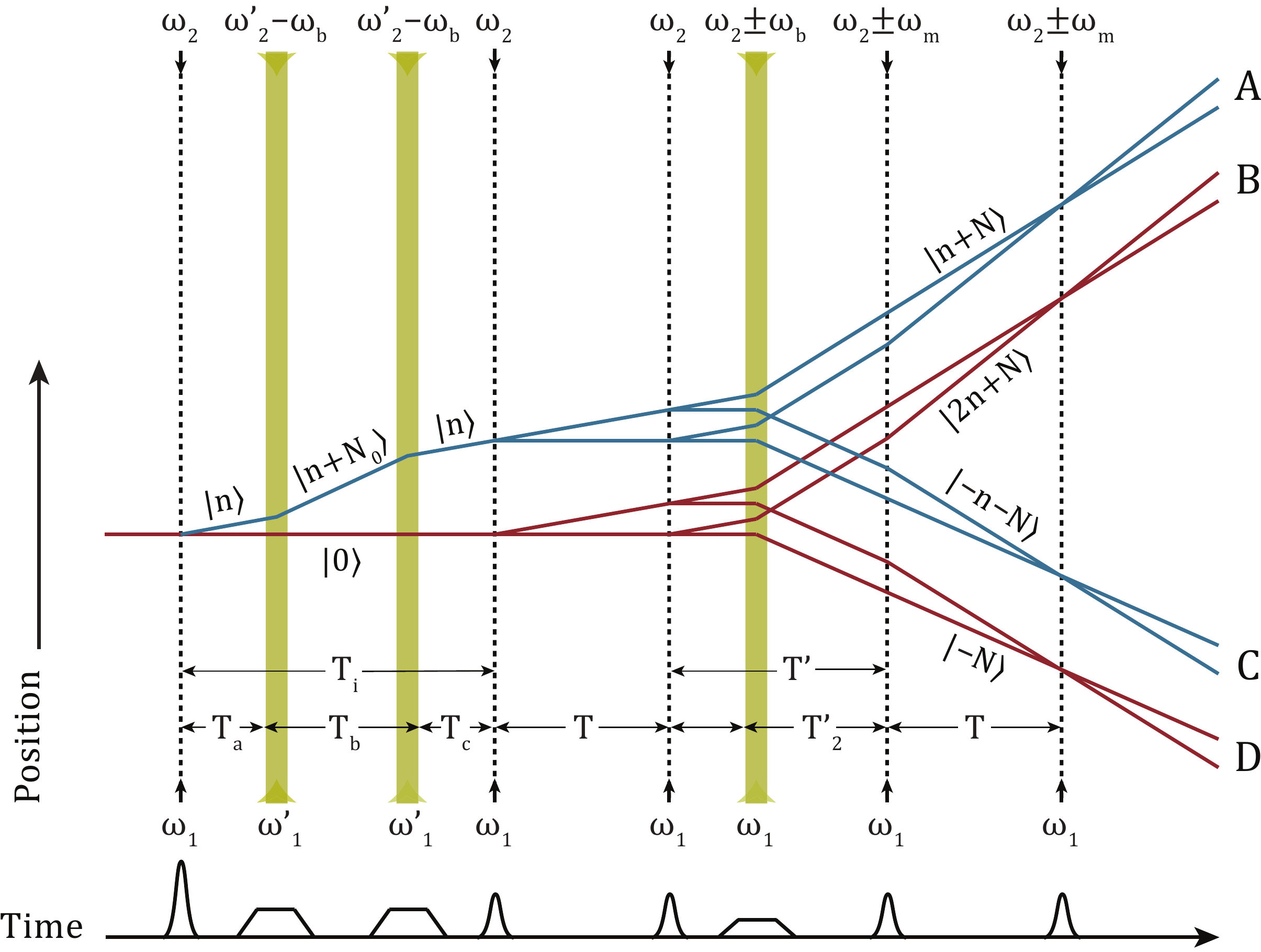}
  \caption{The upper panel shows the pulse sequence and atom trajectories in OSCIs.  Every dashed black line represents a pair of counter-propagating laser pulses that drive Bragg diffraction. Yellow bands represent the optical lattices that drive Bloch oscillations. We split the atom sample into two with one Bragg pulse and two Bloch oscillation sequences, and based on that create two sets of SCIs (the blue one and the red one). The offset between them is controlled by the timing of the two offset-generating Bloch oscillation sequences. The four output ports $A$ to $D$ constitute four output channels. Among these four channels, $AC$ and $BD$ are the regular SCI channels, $BC$ is the $\gamma$-insensitive channel, and $AD$ is the channel most sensitive to $\gamma$. Momentum states are indicated above the lines. In the bottom are the temporal profiles of all the laser pulses. Bragg pulses have Gaussian temporal profiles. 
  Bloch oscillation beams have trapezoid temporal profiles.}
  \label{fig:geometry}
\end{figure}

In this letter, we propose and demonstrate a new scheme: offset simultaneous conjugate interferometers (OSCIs, as shown in Fig.~\ref{fig:geometry}), which can exactly cancel gravity gradients and substantially reduce diffraction phases. We split one atom sample into two using one Bragg diffraction pulse and two Bloch oscillation sequences \cite{AlphaLKB}, and based on that create two sets of SCIs to form a multi-channel interferometer. The vertical offset between the two SCIs can be precisely controlled by the timing of these two Bloch oscillation sequences. We cancel the phase introduced by gravity gradient through properly setting the offset. Importantly, the offset required to cancel the gradient is independent of the gradient itself, and we demonstrate that the undesired diffraction phase from the Bragg beam splitters is also suppressed due to the symmetry of this geometry. Finally, we discuss the potential for using OSCIs to measure the fine structure constant $\alpha$.



For an atom interferometer where the atomic wave packets travel along the lower arm $z_{l}$ and upper arm $z_{u}$, a vertical acceleration gradient $\gamma$ (due to, for example, a gravity gradient, inhomogenous magnetic fields, blackboady radiation, etc.) will introduce a phase shift, to first order in $\gamma$, as
\begin{equation}
\phi_{\gamma}=\frac{m\gamma}{2\hbar}\int_{} \left(z_{l}^2-z_{u}^2\right) dt=\frac{m\gamma A z_{c}}{\hbar},
\end{equation}
where $m$ is the mass of the atom, $\hbar$ is the reduced Planck constant, $A=\int_{} \left(z_{l}-z_{u}\right)dt$ is the space-time area enclosed between the two arms, and $z_{c}= \int_{} \left(z_{l}^{2}-z_{u}^{2}\right)dt / 2A $ is the center of this space-time area. Magnetic fields and blackbody radiation can lead to phase shifts of the same form \cite{AlphaRP, BlackbodyPH}. 

For two simultaneous interferometers whose centers are separated by a vertical distance $\delta z$, the differential phase shift caused by $\gamma$ is
\begin{equation}
\delta\phi_{\gamma}=\frac{m\gamma A}{\hbar}\delta z.
\label{Eq:dphi}
\end{equation}
This $\gamma$-related phase is an important systematic source in many atom interferometer applications \cite{GammaNobili, GammaJason, GTino, AlphaRP}. Several schemes have been proposed to control this effect to sufficient accuracy \cite{GammaChiow, GammaRoura}.

Recently, compensating the gravity gradient has been demonstrated in a Raman-pulse Mach-Zehnder interferometer, which opens up the possibility of an improved determination of the Newton's gravitational constant $G$ and a test of the equivalence principle
\cite{GammaTino,gammaKasevich}.
The key idea is to adjust the frequency of the central $\pi$ pulse to compensate the phase shift produced by $\gamma$. This 
still requires the gravity gradient to be measured, necessarily introducing a corresponding uncertainty.

OSCIs represent an alternate way to cancel the phase shift due to $\gamma$, by overlapping the centers of the two simultaneous interferometers, so that $\delta z =0$ and $\delta\phi_{\gamma}$ vanishes according to Eq.~\ref{Eq:dphi}. Not only does this technique not require a measurement of $\gamma$, it also works for both Bragg and Raman beam splitters, and it can be applied to any atom interferometer geometry, even for the ones that are not intended for a gravity measurement, like  Ramsey-Bord{\'e} interferometers.

Fig.~\ref{fig:geometry} shows the geometry and pulse sequence of OSCIs. We first apply an $n^{th}$ order Bragg pulse, which transfers the momentum of $2n$ photons, and drives the atoms into a superposition of two states $\left|0\right>$ and $\left|n\right>$, where $\left|a\right>$ denotes a momentum eigenstate with momentum $2a\hbar k$ where $k$ is the wavenumber of the laser beam. After a time $T_a$, Bloch oscillations are applied to accelerate the upper arm to the state $\left|n+N_0\right>$. A second deceleration Bloch oscillation sequence then brings this arm back to $\left|n\right>$. As the relative motions of these two arms come from photon recoils, the displacement between them can be precisely controlled by the timing of these two Bloch oscillation beams. 
The initial Bragg pulse and the two Bloch oscillation sequences don't participate in interference. They are referred to as the offset-generating Bragg and Bloch pulses later in this paper.
After another time interval $T_c$, we apply a pair of $n^{th}$ order Bragg beam splitters, a sequence of $N$ Bloch oscillations, and another pair of $n^{th}$ order Bragg beam splitters. Each pair of Bragg beam splitters has a separation time of $T$ and the two pairs are separated by time $T'$. 
The second pair of Bragg beam splitters contains two frequencies, which are shifted by $\pm\omega_m$ relative to the first pair. These beams are referred to as interferometer Bragg and Bloch beams. 
In the end, we have two sets of SCIs (the blue one and the red one) and four output ports ($A$ to $D$), which constitute four output channels. Among these four channels, $AC$ (a channel that outputs the differential phase between $A$ and $C$) and $BD$ are the regular SCI channels. $BC$ is the $\gamma$-insensitive channel, as $B$ and $C$ are spatially overlapped and can be used to cancel the $\gamma$ effect. $AD$, $AC$, and $BD$ are sensitive to $\gamma$, with $AD$ having the largest sensitivity due to it having the largest displacement.  


The diffraction phase is suppressed in the $\gamma$-insensitive channel $BC$, which we can see by examining the symmetries of OSCIs.
After the offset-generating Bragg and Bloch beams, the upper SCI (the blue one) starts from the state $\left|n\right>$, and the lower SCI (the red one) starts from the state $\left|0\right>$. The two interferometers of channel $BC$ now have a symmetric configuration compared to the regular SCI channel $BD$ (or $AC$) where the two interferometers start from the same momentum state. 
This symmetry leads to the cancellation of the diffraction phase from the first two interferometer Bragg pulses.
To show that, we follow the analysis in Ref.~\cite{DiffractionBE, DiffractionRP}. 
We use $\hat B_{n}$ to denote an $n^{th}$ order Bragg pulse. By numerically solving the optical Bloch equations that describe the process of Bragg diffraction \cite{DiffractionRP}, we get the matrix element $\bra{b} \hat B_{n}\ket{a}$ as the complex amplitude for the Bragg pulse to transfer an atom from a momentum state $\ket{a}$ into a momentum state $\ket{b}$. The diffraction phase from the beam splitter $\hat B_{n}$ is thus the argument $\text{arg}(\bra{b} \hat B_{n}\ket{a})$ (here we don't include the laser phase \cite{AIChu}).

Denote the diffraction phase from the $m^{th}$ interferometer Bragg pulse on output port $I$ ($I$=$A$, $B$, $C$ or $D$) as $\phi_{I,m}$. For $A$, the first interferometer Bragg pulse does not change the upper arm's state $\ket{n}$, but drives the lower arm from $\ket{n}$ to $\ket{0}$, thus
$\phi_{A,1}=\text{arg}(\bra{n}\hat{B}_n\ket{n}/\bra{n}\hat{B}_n\ket{0})$. Similarly, we get
\allowdisplaybreaks
\begin{eqnarray}
\phi_{A,1}&=&\phi_{A,2}=\phi_{B,2}=\phi_{C,1}=\text{arg}\left(\frac{\bra{n}\hat{B}_n\ket{n}}{\bra{n}\hat{B}_n\ket{0}}\right), \nonumber\\
\phi_{B,1}&=&\phi_{C,2}=\phi_{D,1}=\phi_{D,2}=\text{arg}\left(\frac{\bra{n}\hat{B}_n\ket{0}}{\bra{0}\hat{B}_n\ket{0}}\right).
\label{Eq:DiffractionPort}
\end{eqnarray}
For every channel $IJ$ ($IJ$=$A$C, $BD$, $BC$ or $AD$), the diffraction phase from the $m^{th}$ interferometer Bragg pulse, denoted as $\phi_{IJ,m}$, is $\phi_{I,m}-\phi_{J,m}$. Using Eq.~\ref{Eq:DiffractionPort}, we get
\allowdisplaybreaks
\begin{eqnarray}
\phi_{AC,1}&=&\phi_{BD,1}=0, \nonumber\\
\phi_{AD,1}&=&-\phi_{BC,1}=\phi_{AC,2}=\phi_{BD,2}=\phi_{AD,2}\nonumber\\
&=&\phi_{BC,2}=\text{arg}\left(\frac{\bra{0}\hat{B}_n\ket{0}\bra{n}\hat{B}_n\ket{n}}{\bra{n}\hat{B}_n\ket{0}^2}\right).
\label{Eq:DiffractionChannel}
\end{eqnarray}
Eq.~\ref{Eq:DiffractionChannel} shows that, 
in channels $AC$ and $BD$, the diffraction phase from the first interferometer Bragg pulse is zero, the main contribution comes from the second one; 
in the $\gamma$-insensitive channel $BC$, the diffraction phases from the first two interferometer Bragg pulses cancel each other; 
on the contrary, in channel $AD$, the first two interferometer Bragg pulses introduce the same diffraction phase. 
Similar analysis shows that the diffraction phase from the third and fourth interferometer Bragg pulses is the same in all channels.
Therefore, the overall diffraction phase is suppressed in channel $BC$ and amplified in channel $AD$ relative to channel $AC$ and $BD$.

Our setup is similar to the one in Ref.~\cite{DiffractionBE}. Cesium atoms are first loaded into a magneto-optical trap and launched upward via moving optical molasses \cite{MolassesPritchard}. The atoms are then further cooled to a few hundred nanokelvin with Raman sideband cooling \cite{RSCChu}. With three subsequent Raman transitions, we prepare cesium atoms into the $\left|F=3, m_F=0\right>$ electronic state with a velocity spread less than 0.1 recoil velocities along the vertical direction. 
Bragg and Bloch beams are then applied during the fight, as show in Fig.~\ref{fig:geometry}. When the atoms fall back down, we apply a resonant detection beam and detect the florescences. The whole sequence is repeated every 2.4 seconds.

We use $5^{th}$ order Bragg pulses ($n=5$), and 125 Bloch oscillations ($N_0=N=125$) for offset-generating and interferometer. All the Bragg and Bloch beams are 15\,GHz blue-detuned from the cesium $\left|F=3\right>\rightarrow\left|F'=4\right>$ D2 transition.
The frequencies of the offset-generating Bloch beams are generated in the same way as the interferometer Bloch beams. So the down-propagating beams contain a pair of frequencies. For them to only interact with the atoms in the momentum state $\left|n\right>$, the frequencies of the counter-propagating beams are decreased by 20\,kHz: $\omega_2-\omega_2'=\omega_1-\omega_1'=2\pi\times20$\,kHz. Atoms left behind by the acceleration Bloch oscillations are mostly in the $\left|n\right>$ state and would spatially overlap with the signals. To suppress these atoms, we apply a velocity-sensitive two-photon Raman pulse to drive the atoms with $2n\hbar k$ momentum from the $\left|F=3\right>$ ground state to the $\left|F=4\right>$ ground state, and then apply a resonant beam to blow away the atoms in the $\left|F=4\right>$ state. These pulses are not shown in Fig.~\ref{fig:geometry}. Atoms left behind by the deceleration Bloch oscillations are in the $\left|n+N\right>$ momentum state and also spatially resolved from the output channels. These atoms do not affect the experiment. 

The main differential phase that we can read out from every channel $IJ$, including the $\gamma$-related term and the diffraction phase, is
\begin{align}
\Phi=-2nT\left[\omega_m -8\omega_r\left(n+N\right)-\omega_r\gamma C_{IJ,T}\right]+ \phi_{IJ},
\label{Eq:phimain}
\end{align}
where  $\omega_r=\hbar k^2/2m_{Cs}$ is the angular recoil frequency of a cesium atom with mass $m_{Cs}$ after scatting a photon with wavenumber $k$. For an arbitrary combination of output ports I and J, $C_{IJ,T}$ is the coefficient of the $\gamma$-related phase, which depends on the channel as well as the pulse separation time $T$:
\allowdisplaybreaks
\begin{eqnarray}
C_{BD,T}&=&\frac{2}{3}n\left(2T^2+3TT'+3T'^2\right)\nonumber \\
&&+\frac{4}{3}N\left(T^2+3TT_2'+3T_2'^2\right),\nonumber \\
C_{AC,T}&=&C_{BD,T},\nonumber \\
C_{AD,T}&=&C_{BD,T}+4\left(nT_i+NT_{b}\right)\left(T+T'\right),\nonumber \\
C_{BC,T}&=&C_{BD,T}-4\left(nT_i+NT_{b}\right)\left(T+T'\right),
\label{Eq:phigamma}
\end{eqnarray}
where $T_i$ is the interval between the offset-generating and first interferometer Bragg pulses, and $T_2'$ is the interval between the interferometer Bloch beam and the third interferometer Bragg pulse. Other high order effects, including the phase terms that depend on the atom velocity, the duration of the Bloch beam, and the gravity acceleration $g$, are not shown for simplicity. 
We fix the timing of the offset-generating Bragg pulse, the acceleration offset-generating Bloch oscillations, the second and the third interferometer Bragg pulses relative to the experiment sequence: $T_{a}=5$\,ms, $T_i+T=155$\,ms, $T'=50$\,ms and $T_2'=45$\,ms. When $T$ changes, we adjust $T_{b}$ according to Eq.~\ref{Eq:phigamma}, so that $C_{BC,T}=0$ and $C_{AD,T}=2C_{BD,T}$. 

%
\begin{figure}[h]
  \centering
  \includegraphics[width=0.4\textwidth]{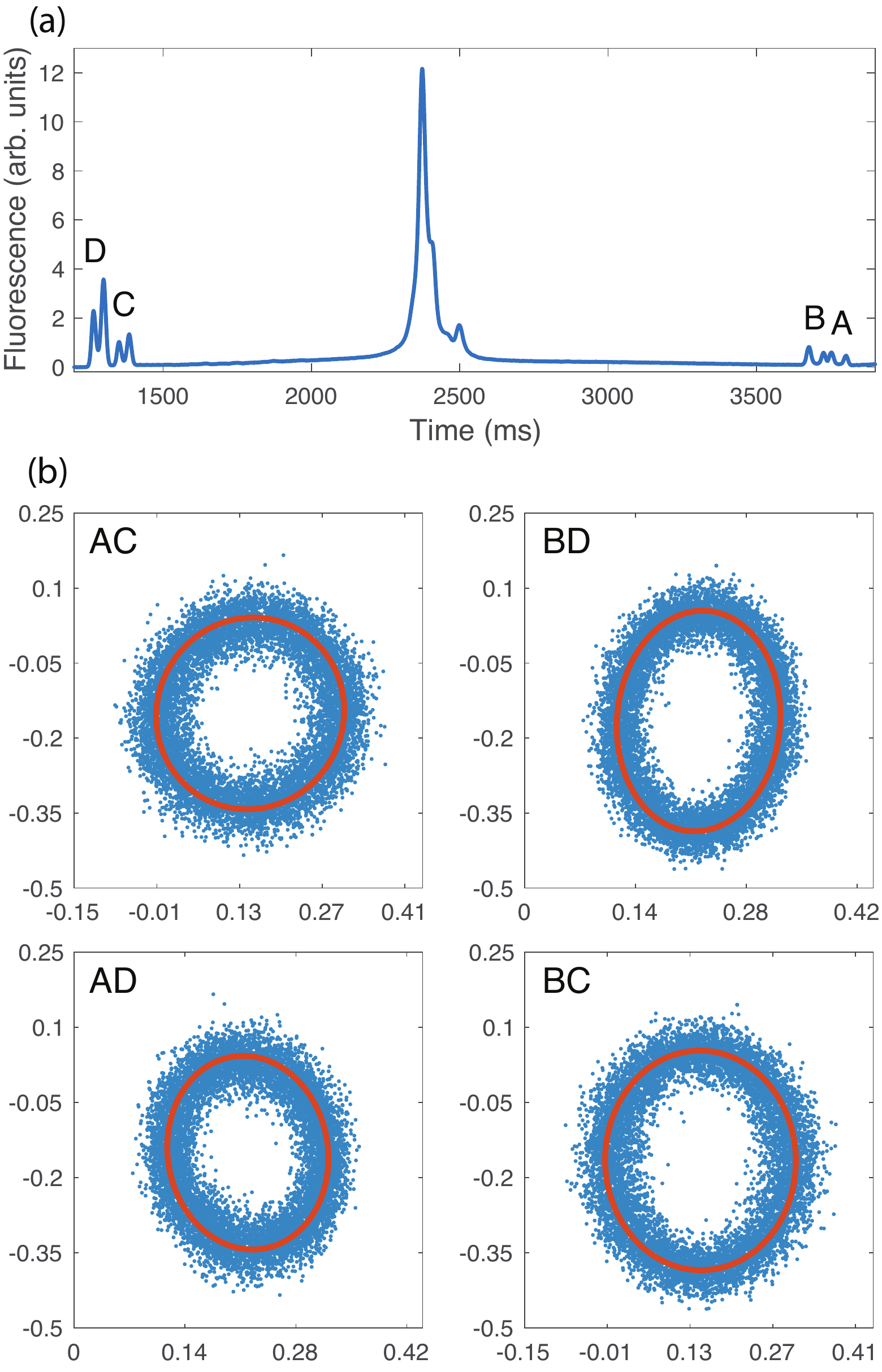}
  \caption{(a) OSCI fluorescence taken at $T = 10$\,ms and $T_{b}=37.6$\,ms. It is an average of 30 measurements. The eight peaks on the two sides correspond to four output ports. The two big peaks in the middle are the atoms not driven by Bloch oscillations, which do not contribute the measurement. (b) Signal from every channel. The x axis is the normalized signal of the lower interferometers ($C$ or $D$), and the y axis is the normalized signal of the upper interferometers ($A$ or $B$). Data was collected over a period of 13 hours. The red curve is the fitted ellipse from every channel.}
  \label{fig:Fluorescence&Ellipses}
\end{figure}


First we set $T$ to 10\,ms. $T_{b}$ is calculated to be 37.6\,ms. This results in an offset of about 3.2\,cm between the two wave packets at the moment of the first interferometer Bragg pulse. Fig.~\ref{fig:Fluorescence&Ellipses}(a) shows the typical fluorescence signal we observe. The eight peaks on the two sides correspond to four output ports. The two peaks in the middle come from the atoms that are not driven by the interferometer Bloch oscillations. These atoms do not contribute to the measurement. We use ellipse fitting \cite{EllipseFisher, EllipseKasevich} to extract the differential phase between two output ports. The ellipses are shown in Fig.~\ref{fig:Fluorescence&Ellipses}(b). The $x$ axis of each ellipse is the normalized signal of the lower interferometer ($C$ or $D$), and the $y$ axis is the normalized signal of the upper interferometer ($A$ or $B$). The $\gamma$-insensitive channel $BC$ has the same $x$ contrast as channel $AC$ (about 16\%), and the same $y$ contrast as channel $BD$ (about 22\%), demonstrating that introducing an offset doesn't result in decoherence.  After 13 hours of integration, we reach an uncertainty of about 3 part-per-billion (ppb) in the differential phase for every channel, which corresponds to 1.5\,ppb statistical uncertainty in $\alpha$ \cite{AlphaRP}. This demonstrates the world-class sensitivity of OSCIs.
 
The sensitivity is currently limited by the low signal-to-noise ratio (SNR), due to two effects. First, single photon scattering introduced by the two offset-generating Bloch oscillation sequences reduce the total signal size. 
Second, the laser beams have a $1/e^2$ waist of about 3.4\,mm, which is comparable to the size of the atomic sample. The atoms away from the center of the cloud experience low pulse intensities. This lowers the efficiency of the laser pulses. 
Increasing the detuning of the laser pulses to suppress single photon scattering, and using broader laser beams to drive Bragg diffraction and Bloch oscillations (so the intensity is more uniform across the atomic sample), is expected to improve the SNR.

\begin{figure}[ht]
  \centering
  \includegraphics[width=0.47\textwidth]{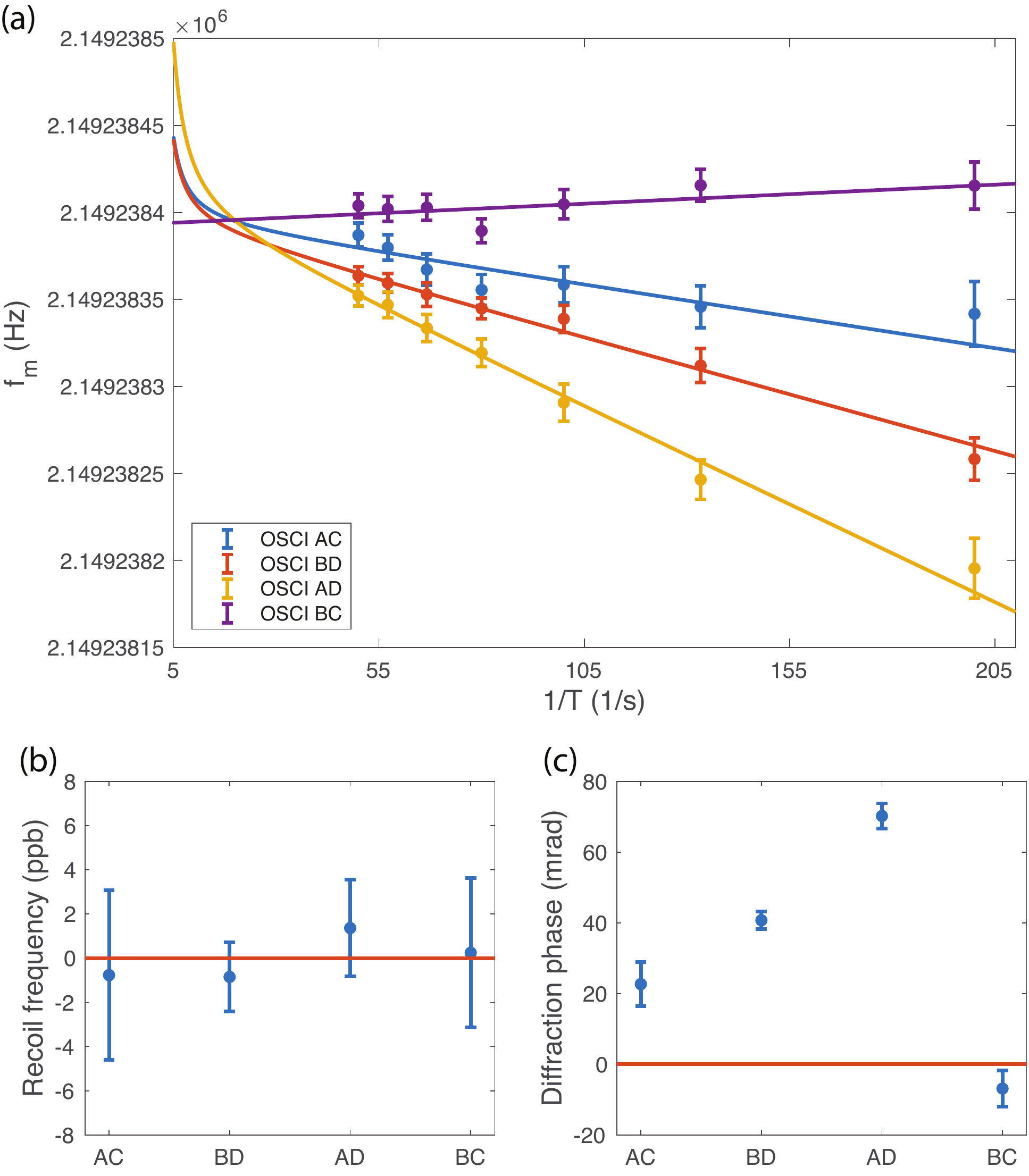}
  \caption{(a) Measured frequency vs. 1/$T$. The data points are from experimental measurements and curves are corresponding fittings using the functional form of Eq.~\ref{eq:fitfmeasured}. The $\gamma$ value comes from Ref.~\cite{AlphaRP}. Channel $BC$ has no $\gamma$ effect, the fitting is a straight line. Data was taken over two successive days. (b) Fitted recoil frequencies in all channels. They are consistent within in 1$\,\sigma$ uncertainty. The red line is the average of these frequencies. (c) Fitted diffraction phases from all channels. The red line indicates zero. A 6-fold suppression of the diffraction phase in channel $BC$ was observed compared to channel $BD$.}
  \label{fig:MeasuredFrequency}
\end{figure}


Next, we show the suppression of the diffraction phase and the consistency of the recoil frequency from every channel. We vary the pulse separation time $T$ from 5\,ms to 20\,ms, and adjust $T_{b}$ accordingly. 
At each $T$, we adjust $\omega_m$ to the point where the total phase $\Phi$ is zero. We define the measured frequency as $f_{m}=\omega_m/(2\pi)$.
According Eq.~\ref{Eq:phimain} and Eq.~\ref{Eq:phigamma},  $f_{m}$ can be fitted as a function of $1/T$ with two fitting parameters, the diffraction phase and the recoil frequency $f_r=\omega_r/(2\pi)$, 
\begin{equation}
f_{m}=\frac{\phi_{IJ}}{4n\pi T}+\left[8(n+N)+\gamma\times C_{IJ,T}\right]f_{r}.
\label{eq:fitfmeasured}
\end{equation}

Fig.~\ref{fig:MeasuredFrequency}(a) shows the measurement results and the corresponding fittings. 
At short $T$s, the contribution from the gravity gradient is small. $f_m$ is roughly linear in $1/T$, with a slope proportional to the diffraction phase. The coefficients $C_{IJ,T}$ scale with $T^2$. As $T$ increases, the effect from the gravity gradient becomes more pronounced, thus the fit curve is no longer linear. This represents a potentially large systemic effect in the $\alpha$ measurement. For instance, at $T=80$\,ms, the gravity gradient will shift the total phase in channel $BD$ by 8\,ppb, which is over two orders of magnitude larger than the required accuracy of the recoil frequency. By properly setting the timing, sensitivity to $\gamma$ in channel $BC$ is cancelled. The fitting for this channel is a straight line.


Fig.~\ref{fig:MeasuredFrequency}(b) shows the fitted recoil frequency from all channels. They agree with each other within 1\,$\sigma$ uncertainty. This demonstrates the consistency of the output from every channel. 
Fig. \ref{fig:MeasuredFrequency}(c) shows the fitted diffraction phase from all channels: $\phi_{AC}=22\pm6\text{\,mrad}$, $\phi_{BD}=41\pm3\text{\,mrad}$, $\phi_{AD}=70\pm4\text{\,mrad}$ and $\phi_{BC}=-7\pm5\text{\,mrad}$. The diffraction phase is suppressed by a factor of 6 in channel $BC$ compare to channel $BD$, and a factor of 10 compare to channel $AD$. 



A measurement of $\alpha$ using atom interferometers has two leading systematic sources: spatial variations of acceleration and the non-Gaussian wavefronts \cite{AlphaRP}. With this new OSCI scheme, the uncertainty from the acceleration gradient will now be negligible; the systematic uncertainty from higher-order variations is estimated to be 0.005\,ppb (based on modeling mass distributions near the vacuum system), yielding a 4-fold reduction in the systematic error. The effects from non-Gaussian wavefronts can be suppressed by driving Bragg diffraction and Bloch oscillations with larger-width laser beams, which as mentioned above also improves the sensitivity. Combing with other upgrades described in \cite{Chenghui} that will increase statistical sensitivity, we expect to improve the accuracy of $\alpha$ to ten part-per-trillion level, which will provide insights in the test of Standard Model.

The ability to control the offset in OSCIs also enables new ways to check many systematics. For example, with the increased vertical separation between $A$ and $D$, OSCIs can measure $\gamma$ using channel $AD$. Channel $AC$ and $BD$ are two SCIs operated simultaneously at different heights. Comparing the results from these two channels allows us to identify and reduce effects such as the inhomogeneous magnetic field along the vertical direction, the divergence of the laser beams and the stray light reflected by the vacuum chamber \cite{AlphaRP}. Because these two channels have the same phase produced by gravity gradient, a comparison between them will also place a bound on third order gravity variation (the gradient of the gravity gradient). These systematic checks can be done simultaneously with data taken for a measurement of the recoil frequency.


\begin{thebibliography}{33}%
\makeatletter
\providecommand \@ifxundefined [1]{%
 \@ifx{#1\undefined}
}%
\providecommand \@ifnum [1]{%
 \ifnum #1\expandafter \@firstoftwo
 \else \expandafter \@secondoftwo
 \fi
}%
\providecommand \@ifx [1]{%
 \ifx #1\expandafter \@firstoftwo
 \else \expandafter \@secondoftwo
 \fi
}%
\providecommand \natexlab [1]{#1}%
\providecommand \enquote  [1]{``#1''}%
\providecommand \bibnamefont  [1]{#1}%
\providecommand \bibfnamefont [1]{#1}%
\providecommand \citenamefont [1]{#1}%
\providecommand \href@noop [0]{\@secondoftwo}%
\providecommand \href [0]{\begingroup \@sanitize@url \@href}%
\providecommand \@href[1]{\@@startlink{#1}\@@href}%
\providecommand \@@href[1]{\endgroup#1\@@endlink}%
\providecommand \@sanitize@url [0]{\catcode `\\12\catcode `\$12\catcode
  `\&12\catcode `\#12\catcode `\^12\catcode `\_12\catcode `\%12\relax}%
\providecommand \@@startlink[1]{}%
\providecommand \@@endlink[0]{}%
\providecommand \url  [0]{\begingroup\@sanitize@url \@url }%
\providecommand \@url [1]{\endgroup\@href {#1}{\urlprefix }}%
\providecommand \urlprefix  [0]{URL }%
\providecommand \Eprint [0]{\href }%
\providecommand \doibase [0]{http://dx.doi.org/}%
\providecommand \selectlanguage [0]{\@gobble}%
\providecommand \bibinfo  [0]{\@secondoftwo}%
\providecommand \bibfield  [0]{\@secondoftwo}%
\providecommand \translation [1]{[#1]}%
\providecommand \BibitemOpen [0]{}%
\providecommand \bibitemStop [0]{}%
\providecommand \bibitemNoStop [0]{.\EOS\space}%
\providecommand \EOS [0]{\spacefactor3000\relax}%
\providecommand \BibitemShut  [1]{\csname bibitem#1\endcsname}%
\let\auto@bib@innerbib\@empty
\bibitem [{\citenamefont {Peters}\ \emph {et~al.}(1999)\citenamefont {Peters},
  \citenamefont {Chung},\ and\ \citenamefont {Chu}}]{InertialChu}%
  \BibitemOpen
  \bibfield  {author} {\bibinfo {author} {\bibfnamefont {A.}~\bibnamefont
  {Peters}}, \bibinfo {author} {\bibfnamefont {K.~Y.}\ \bibnamefont {Chung}}, \
  and\ \bibinfo {author} {\bibfnamefont {S.}~\bibnamefont {Chu}},\ }\href
  {\doibase 10.1038/23655} {\bibfield  {journal} {\bibinfo  {journal} {Nature}\
  }\textbf {\bibinfo {volume} {400}},\ \bibinfo {pages} {849} (\bibinfo {year}
  {1999})}\BibitemShut {NoStop}%
\bibitem [{\citenamefont {Geiger}\ \emph {et~al.}(2011)\citenamefont {Geiger},
  \citenamefont {M\'enoret}, \citenamefont {Stern}, \citenamefont {Zahzam},
  \citenamefont {Cheinet}, \citenamefont {Battelier}, \citenamefont {Villing},
  \citenamefont {Moron}, \citenamefont {Lours}, \citenamefont {Bidel},
  \citenamefont {Bresson}, \citenamefont {Landragin},\ and\ \citenamefont
  {Bouyer}}]{InertialBouyer}%
  \BibitemOpen
  \bibfield  {author} {\bibinfo {author} {\bibfnamefont {R.}~\bibnamefont
  {Geiger}}, \bibinfo {author} {\bibfnamefont {V.}~\bibnamefont {M\'enoret}},
  \bibinfo {author} {\bibfnamefont {G.}~\bibnamefont {Stern}}, \bibinfo
  {author} {\bibfnamefont {N.}~\bibnamefont {Zahzam}}, \bibinfo {author}
  {\bibfnamefont {P.}~\bibnamefont {Cheinet}}, \bibinfo {author} {\bibfnamefont
  {B.}~\bibnamefont {Battelier}}, \bibinfo {author} {\bibfnamefont
  {A.}~\bibnamefont {Villing}}, \bibinfo {author} {\bibfnamefont
  {F.}~\bibnamefont {Moron}}, \bibinfo {author} {\bibfnamefont
  {M.}~\bibnamefont {Lours}}, \bibinfo {author} {\bibfnamefont
  {Y.}~\bibnamefont {Bidel}}, \bibinfo {author} {\bibfnamefont
  {A.}~\bibnamefont {Bresson}}, \bibinfo {author} {\bibfnamefont
  {A.}~\bibnamefont {Landragin}}, \ and\ \bibinfo {author} {\bibfnamefont
  {P.}~\bibnamefont {Bouyer}},\ }\href {\doibase 10.1038/ncomms1479} {\bibfield
   {journal} {\bibinfo  {journal} {Nature Communications}\ }\textbf {\bibinfo
  {volume} {2}},\ \bibinfo {pages} {474} (\bibinfo {year} {2011})}\BibitemShut
  {NoStop}%
\bibitem [{\citenamefont {Sugarbaker}\ \emph {et~al.}(2013)\citenamefont
  {Sugarbaker}, \citenamefont {Dickerson}, \citenamefont {Hogan}, \citenamefont
  {Johnson},\ and\ \citenamefont {Kasevich}}]{InertialKasevich}%
  \BibitemOpen
  \bibfield  {author} {\bibinfo {author} {\bibfnamefont {A.}~\bibnamefont
  {Sugarbaker}}, \bibinfo {author} {\bibfnamefont {S.~M.}\ \bibnamefont
  {Dickerson}}, \bibinfo {author} {\bibfnamefont {J.~M.}\ \bibnamefont
  {Hogan}}, \bibinfo {author} {\bibfnamefont {D.~M.~S.}\ \bibnamefont
  {Johnson}}, \ and\ \bibinfo {author} {\bibfnamefont {M.~A.}\ \bibnamefont
  {Kasevich}},\ }\href {\doibase 10.1103/PhysRevLett.111.113002} {\bibfield
  {journal} {\bibinfo  {journal} {Phys. Rev. Lett.}\ }\textbf {\bibinfo
  {volume} {111}},\ \bibinfo {pages} {113002} (\bibinfo {year}
  {2013})}\BibitemShut {NoStop}%
\bibitem [{\citenamefont {Dutta}\ \emph {et~al.}(2016)\citenamefont {Dutta},
  \citenamefont {Savoie}, \citenamefont {Fang}, \citenamefont {Venon},
  \citenamefont {Garrido~Alzar}, \citenamefont {Geiger},\ and\ \citenamefont
  {Landragin}}]{InertialLandragin}%
  \BibitemOpen
  \bibfield  {author} {\bibinfo {author} {\bibfnamefont {I.}~\bibnamefont
  {Dutta}}, \bibinfo {author} {\bibfnamefont {D.}~\bibnamefont {Savoie}},
  \bibinfo {author} {\bibfnamefont {B.}~\bibnamefont {Fang}}, \bibinfo {author}
  {\bibfnamefont {B.}~\bibnamefont {Venon}}, \bibinfo {author} {\bibfnamefont
  {C.~L.}\ \bibnamefont {Garrido~Alzar}}, \bibinfo {author} {\bibfnamefont
  {R.}~\bibnamefont {Geiger}}, \ and\ \bibinfo {author} {\bibfnamefont
  {A.}~\bibnamefont {Landragin}},\ }\href {\doibase
  10.1103/PhysRevLett.116.183003} {\bibfield  {journal} {\bibinfo  {journal}
  {Phys. Rev. Lett.}\ }\textbf {\bibinfo {volume} {116}},\ \bibinfo {pages}
  {183003} (\bibinfo {year} {2016})}\BibitemShut {NoStop}%
\bibitem [{\citenamefont {Mingjie~Xin}(2018)}]{InertialShauYu}%
  \BibitemOpen
  \bibfield  {author} {\bibinfo {author} {\bibfnamefont {Z.~C. S.-Y.~L.}\
  \bibnamefont {Mingjie~Xin}, \bibfnamefont {Wui Seng~Leong}},\ }\href
  {\doibase 10.1126/sciadv.1701723} {\bibfield  {journal} {\bibinfo  {journal}
  {Science Advances}\ }\textbf {\bibinfo {volume} {4}} (\bibinfo {year}
  {2018}),\ 10.1126/sciadv.1701723}\BibitemShut {NoStop}%
\bibitem [{\citenamefont {Fray}\ \emph {et~al.}(2004)\citenamefont {Fray},
  \citenamefont {Diez}, \citenamefont {H\"ansch},\ and\ \citenamefont
  {Weitz}}]{EEPWeitz}%
  \BibitemOpen
  \bibfield  {author} {\bibinfo {author} {\bibfnamefont {S.}~\bibnamefont
  {Fray}}, \bibinfo {author} {\bibfnamefont {C.~A.}\ \bibnamefont {Diez}},
  \bibinfo {author} {\bibfnamefont {T.~W.}\ \bibnamefont {H\"ansch}}, \ and\
  \bibinfo {author} {\bibfnamefont {M.}~\bibnamefont {Weitz}},\ }\href
  {\doibase 10.1103/PhysRevLett.93.240404} {\bibfield  {journal} {\bibinfo
  {journal} {Phys. Rev. Lett.}\ }\textbf {\bibinfo {volume} {93}},\ \bibinfo
  {pages} {240404} (\bibinfo {year} {2004})}\BibitemShut {NoStop}%
\bibitem [{\citenamefont {Schlippert}\ \emph {et~al.}(2014)\citenamefont
  {Schlippert}, \citenamefont {Hartwig}, \citenamefont {Albers}, \citenamefont
  {Richardson}, \citenamefont {Schubert}, \citenamefont {Roura}, \citenamefont
  {Schleich}, \citenamefont {Ertmer},\ and\ \citenamefont {Rasel}}]{EEPRasel}%
  \BibitemOpen
  \bibfield  {author} {\bibinfo {author} {\bibfnamefont {D.}~\bibnamefont
  {Schlippert}}, \bibinfo {author} {\bibfnamefont {J.}~\bibnamefont {Hartwig}},
  \bibinfo {author} {\bibfnamefont {H.}~\bibnamefont {Albers}}, \bibinfo
  {author} {\bibfnamefont {L.~L.}\ \bibnamefont {Richardson}}, \bibinfo
  {author} {\bibfnamefont {C.}~\bibnamefont {Schubert}}, \bibinfo {author}
  {\bibfnamefont {A.}~\bibnamefont {Roura}}, \bibinfo {author} {\bibfnamefont
  {W.~P.}\ \bibnamefont {Schleich}}, \bibinfo {author} {\bibfnamefont
  {W.}~\bibnamefont {Ertmer}}, \ and\ \bibinfo {author} {\bibfnamefont {E.~M.}\
  \bibnamefont {Rasel}},\ }\href {\doibase 10.1103/PhysRevLett.112.203002}
  {\bibfield  {journal} {\bibinfo  {journal} {Phys. Rev. Lett.}\ }\textbf
  {\bibinfo {volume} {112}},\ \bibinfo {pages} {203002} (\bibinfo {year}
  {2014})}\BibitemShut {NoStop}%
\bibitem [{\citenamefont {Tarallo}\ \emph {et~al.}(2014)\citenamefont
  {Tarallo}, \citenamefont {Mazzoni}, \citenamefont {Poli}, \citenamefont
  {Sutyrin}, \citenamefont {Zhang},\ and\ \citenamefont {Tino}}]{EEPTino}%
  \BibitemOpen
  \bibfield  {author} {\bibinfo {author} {\bibfnamefont {M.~G.}\ \bibnamefont
  {Tarallo}}, \bibinfo {author} {\bibfnamefont {T.}~\bibnamefont {Mazzoni}},
  \bibinfo {author} {\bibfnamefont {N.}~\bibnamefont {Poli}}, \bibinfo {author}
  {\bibfnamefont {D.~V.}\ \bibnamefont {Sutyrin}}, \bibinfo {author}
  {\bibfnamefont {X.}~\bibnamefont {Zhang}}, \ and\ \bibinfo {author}
  {\bibfnamefont {G.~M.}\ \bibnamefont {Tino}},\ }\href {\doibase
  10.1103/PhysRevLett.113.023005} {\bibfield  {journal} {\bibinfo  {journal}
  {Phys. Rev. Lett.}\ }\textbf {\bibinfo {volume} {113}},\ \bibinfo {pages}
  {023005} (\bibinfo {year} {2014})}\BibitemShut {NoStop}%
\bibitem [{\citenamefont {Zhou}\ \emph {et~al.}(2015)\citenamefont {Zhou},
  \citenamefont {Long}, \citenamefont {Tang}, \citenamefont {Chen},
  \citenamefont {Gao}, \citenamefont {Peng}, \citenamefont {Duan},
  \citenamefont {Zhong}, \citenamefont {Xiong}, \citenamefont {Wang},
  \citenamefont {Zhang},\ and\ \citenamefont {Zhan}}]{EEPZhan}%
  \BibitemOpen
  \bibfield  {author} {\bibinfo {author} {\bibfnamefont {L.}~\bibnamefont
  {Zhou}}, \bibinfo {author} {\bibfnamefont {S.}~\bibnamefont {Long}}, \bibinfo
  {author} {\bibfnamefont {B.}~\bibnamefont {Tang}}, \bibinfo {author}
  {\bibfnamefont {X.}~\bibnamefont {Chen}}, \bibinfo {author} {\bibfnamefont
  {F.}~\bibnamefont {Gao}}, \bibinfo {author} {\bibfnamefont {W.}~\bibnamefont
  {Peng}}, \bibinfo {author} {\bibfnamefont {W.}~\bibnamefont {Duan}}, \bibinfo
  {author} {\bibfnamefont {J.}~\bibnamefont {Zhong}}, \bibinfo {author}
  {\bibfnamefont {Z.}~\bibnamefont {Xiong}}, \bibinfo {author} {\bibfnamefont
  {J.}~\bibnamefont {Wang}}, \bibinfo {author} {\bibfnamefont {Y.}~\bibnamefont
  {Zhang}}, \ and\ \bibinfo {author} {\bibfnamefont {M.}~\bibnamefont {Zhan}},\
  }\href {\doibase 10.1103/PhysRevLett.115.013004} {\bibfield  {journal}
  {\bibinfo  {journal} {Phys. Rev. Lett.}\ }\textbf {\bibinfo {volume} {115}},\
  \bibinfo {pages} {013004} (\bibinfo {year} {2015})}\BibitemShut {NoStop}%
\bibitem [{\citenamefont {Duan}\ \emph {et~al.}(2016)\citenamefont {Duan},
  \citenamefont {Deng}, \citenamefont {Zhou}, \citenamefont {Zhang},
  \citenamefont {Xu}, \citenamefont {Xiong}, \citenamefont {Xu}, \citenamefont
  {Shao}, \citenamefont {Luo},\ and\ \citenamefont {Hu}}]{EEPHu}%
  \BibitemOpen
  \bibfield  {author} {\bibinfo {author} {\bibfnamefont {X.-C.}\ \bibnamefont
  {Duan}}, \bibinfo {author} {\bibfnamefont {X.-B.}\ \bibnamefont {Deng}},
  \bibinfo {author} {\bibfnamefont {M.-K.}\ \bibnamefont {Zhou}}, \bibinfo
  {author} {\bibfnamefont {K.}~\bibnamefont {Zhang}}, \bibinfo {author}
  {\bibfnamefont {W.-J.}\ \bibnamefont {Xu}}, \bibinfo {author} {\bibfnamefont
  {F.}~\bibnamefont {Xiong}}, \bibinfo {author} {\bibfnamefont {Y.-Y.}\
  \bibnamefont {Xu}}, \bibinfo {author} {\bibfnamefont {C.-G.}\ \bibnamefont
  {Shao}}, \bibinfo {author} {\bibfnamefont {J.}~\bibnamefont {Luo}}, \ and\
  \bibinfo {author} {\bibfnamefont {Z.-K.}\ \bibnamefont {Hu}},\ }\href
  {\doibase 10.1103/PhysRevLett.117.023001} {\bibfield  {journal} {\bibinfo
  {journal} {Phys. Rev. Lett.}\ }\textbf {\bibinfo {volume} {117}},\ \bibinfo
  {pages} {023001} (\bibinfo {year} {2016})}\BibitemShut {NoStop}%
\bibitem [{\citenamefont {Fixler}\ \emph {et~al.}(2007)\citenamefont {Fixler},
  \citenamefont {Foster}, \citenamefont {McGuirk},\ and\ \citenamefont
  {Kasevich}}]{GKasevich}%
  \BibitemOpen
  \bibfield  {author} {\bibinfo {author} {\bibfnamefont {J.~B.}\ \bibnamefont
  {Fixler}}, \bibinfo {author} {\bibfnamefont {G.~T.}\ \bibnamefont {Foster}},
  \bibinfo {author} {\bibfnamefont {J.~M.}\ \bibnamefont {McGuirk}}, \ and\
  \bibinfo {author} {\bibfnamefont {M.~A.}\ \bibnamefont {Kasevich}},\ }\href
  {\doibase 10.1126/science.1135459} {\bibfield  {journal} {\bibinfo  {journal}
  {Science}\ }\textbf {\bibinfo {volume} {315}},\ \bibinfo {pages} {74}
  (\bibinfo {year} {2007})}\BibitemShut {NoStop}%
\bibitem [{\citenamefont {Rosi}\ \emph {et~al.}(2014)\citenamefont {Rosi},
  \citenamefont {Sorrentino}, \citenamefont {Cacciapuoti}, \citenamefont
  {Prevedelli},\ and\ \citenamefont {Tino}}]{GTino}%
  \BibitemOpen
  \bibfield  {author} {\bibinfo {author} {\bibfnamefont {G.}~\bibnamefont
  {Rosi}}, \bibinfo {author} {\bibfnamefont {F.}~\bibnamefont {Sorrentino}},
  \bibinfo {author} {\bibfnamefont {L.}~\bibnamefont {Cacciapuoti}}, \bibinfo
  {author} {\bibfnamefont {M.}~\bibnamefont {Prevedelli}}, \ and\ \bibinfo
  {author} {\bibfnamefont {G.~M.}\ \bibnamefont {Tino}},\ }\href {\doibase
  10.1038/nature13433} {\bibfield  {journal} {\bibinfo  {journal} {Nature}\
  }\textbf {\bibinfo {volume} {510}},\ \bibinfo {pages} {518} (\bibinfo {year}
  {2014})}\BibitemShut {NoStop}%
\bibitem [{\citenamefont {Elder}\ \emph {et~al.}(2016)\citenamefont {Elder},
  \citenamefont {Khoury}, \citenamefont {Haslinger}, \citenamefont {Jaffe},
  \citenamefont {M\"uller},\ and\ \citenamefont {Hamilton}}]{ChameleonBE}%
  \BibitemOpen
  \bibfield  {author} {\bibinfo {author} {\bibfnamefont {B.}~\bibnamefont
  {Elder}}, \bibinfo {author} {\bibfnamefont {J.}~\bibnamefont {Khoury}},
  \bibinfo {author} {\bibfnamefont {P.}~\bibnamefont {Haslinger}}, \bibinfo
  {author} {\bibfnamefont {M.}~\bibnamefont {Jaffe}}, \bibinfo {author}
  {\bibfnamefont {H.}~\bibnamefont {M\"uller}}, \ and\ \bibinfo {author}
  {\bibfnamefont {P.}~\bibnamefont {Hamilton}},\ }\href {\doibase
  10.1103/PhysRevD.94.044051} {\bibfield  {journal} {\bibinfo  {journal} {Phys.
  Rev. D}\ }\textbf {\bibinfo {volume} {94}},\ \bibinfo {pages} {044051}
  (\bibinfo {year} {2016})}\BibitemShut {NoStop}%
\bibitem [{\citenamefont {Hamilton}\ \emph {et~al.}(2015)\citenamefont
  {Hamilton}, \citenamefont {Jaffe}, \citenamefont {Haslinger}, \citenamefont
  {Simmons}, \citenamefont {M{\"u}ller},\ and\ \citenamefont
  {Khoury}}]{DarkEnergyPH}%
  \BibitemOpen
  \bibfield  {author} {\bibinfo {author} {\bibfnamefont {P.}~\bibnamefont
  {Hamilton}}, \bibinfo {author} {\bibfnamefont {M.}~\bibnamefont {Jaffe}},
  \bibinfo {author} {\bibfnamefont {P.}~\bibnamefont {Haslinger}}, \bibinfo
  {author} {\bibfnamefont {Q.}~\bibnamefont {Simmons}}, \bibinfo {author}
  {\bibfnamefont {H.}~\bibnamefont {M{\"u}ller}}, \ and\ \bibinfo {author}
  {\bibfnamefont {J.}~\bibnamefont {Khoury}},\ }\href {\doibase
  10.1126/science.aaa8883} {\bibfield  {journal} {\bibinfo  {journal}
  {Science}\ }\textbf {\bibinfo {volume} {349}},\ \bibinfo {pages} {849}
  (\bibinfo {year} {2015})}\BibitemShut {NoStop}%
\bibitem [{\citenamefont {D'Amico}\ \emph {et~al.}(2017)\citenamefont
  {D'Amico}, \citenamefont {Rosi}, \citenamefont {Zhan}, \citenamefont
  {Cacciapuoti}, \citenamefont {Fattori},\ and\ \citenamefont
  {Tino}}]{GammaTino}%
  \BibitemOpen
  \bibfield  {author} {\bibinfo {author} {\bibfnamefont {G.}~\bibnamefont
  {D'Amico}}, \bibinfo {author} {\bibfnamefont {G.}~\bibnamefont {Rosi}},
  \bibinfo {author} {\bibfnamefont {S.}~\bibnamefont {Zhan}}, \bibinfo {author}
  {\bibfnamefont {L.}~\bibnamefont {Cacciapuoti}}, \bibinfo {author}
  {\bibfnamefont {M.}~\bibnamefont {Fattori}}, \ and\ \bibinfo {author}
  {\bibfnamefont {G.~M.}\ \bibnamefont {Tino}},\ }\href {\doibase
  10.1103/PhysRevLett.119.253201} {\bibfield  {journal} {\bibinfo  {journal}
  {Phys. Rev. Lett.}\ }\textbf {\bibinfo {volume} {119}},\ \bibinfo {pages}
  {253201} (\bibinfo {year} {2017})}\BibitemShut {NoStop}%
\bibitem [{\citenamefont {Parker}\ \emph {et~al.}(2018)\citenamefont {Parker},
  \citenamefont {Yu}, \citenamefont {Zhong}, \citenamefont {Estey},\ and\
  \citenamefont {M{\"u}ller}}]{AlphaRP}%
  \BibitemOpen
  \bibfield  {author} {\bibinfo {author} {\bibfnamefont {R.~H.}\ \bibnamefont
  {Parker}}, \bibinfo {author} {\bibfnamefont {C.}~\bibnamefont {Yu}}, \bibinfo
  {author} {\bibfnamefont {W.}~\bibnamefont {Zhong}}, \bibinfo {author}
  {\bibfnamefont {B.}~\bibnamefont {Estey}}, \ and\ \bibinfo {author}
  {\bibfnamefont {H.}~\bibnamefont {M{\"u}ller}},\ }\href {\doibase
  10.1126/science.aap7706} {\bibfield  {journal} {\bibinfo  {journal}
  {Science}\ }\textbf {\bibinfo {volume} {360}},\ \bibinfo {pages} {191}
  (\bibinfo {year} {2018})}\BibitemShut {NoStop}%
\bibitem [{\citenamefont {Chiow}\ \emph {et~al.}(2011)\citenamefont {Chiow},
  \citenamefont {Kovachy}, \citenamefont {Chien},\ and\ \citenamefont
  {Kasevich}}]{BraggKasevich}%
  \BibitemOpen
  \bibfield  {author} {\bibinfo {author} {\bibfnamefont {S.-w.}\ \bibnamefont
  {Chiow}}, \bibinfo {author} {\bibfnamefont {T.}~\bibnamefont {Kovachy}},
  \bibinfo {author} {\bibfnamefont {H.-C.}\ \bibnamefont {Chien}}, \ and\
  \bibinfo {author} {\bibfnamefont {M.~A.}\ \bibnamefont {Kasevich}},\ }\href
  {\doibase 10.1103/PhysRevLett.107.130403} {\bibfield  {journal} {\bibinfo
  {journal} {Phys. Rev. Lett.}\ }\textbf {\bibinfo {volume} {107}},\ \bibinfo
  {pages} {130403} (\bibinfo {year} {2011})}\BibitemShut {NoStop}%
\bibitem [{\citenamefont {M\"uller}\ \emph {et~al.}(2008)\citenamefont
  {M\"uller}, \citenamefont {Chiow},\ and\ \citenamefont {Chu}}]{BraggChu}%
  \BibitemOpen
  \bibfield  {author} {\bibinfo {author} {\bibfnamefont {H.}~\bibnamefont
  {M\"uller}}, \bibinfo {author} {\bibfnamefont {S.-w.}\ \bibnamefont {Chiow}},
  \ and\ \bibinfo {author} {\bibfnamefont {S.}~\bibnamefont {Chu}},\ }\href
  {\doibase 10.1103/PhysRevA.77.023609} {\bibfield  {journal} {\bibinfo
  {journal} {Phys. Rev. A}\ }\textbf {\bibinfo {volume} {77}},\ \bibinfo
  {pages} {023609} (\bibinfo {year} {2008})}\BibitemShut {NoStop}%
\bibitem [{\citenamefont {Bouchendira}\ \emph {et~al.}(2011)\citenamefont
  {Bouchendira}, \citenamefont {Clad\'e}, \citenamefont {Guellati-Kh\'elifa},
  \citenamefont {Nez},\ and\ \citenamefont {Biraben}}]{AlphaLKB}%
  \BibitemOpen
  \bibfield  {author} {\bibinfo {author} {\bibfnamefont {R.}~\bibnamefont
  {Bouchendira}}, \bibinfo {author} {\bibfnamefont {P.}~\bibnamefont
  {Clad\'e}}, \bibinfo {author} {\bibfnamefont {S.}~\bibnamefont
  {Guellati-Kh\'elifa}}, \bibinfo {author} {\bibfnamefont {F.}~\bibnamefont
  {Nez}}, \ and\ \bibinfo {author} {\bibfnamefont {F.}~\bibnamefont
  {Biraben}},\ }\href {\doibase 10.1103/PhysRevLett.106.080801} {\bibfield
  {journal} {\bibinfo  {journal} {Phys. Rev. Lett.}\ }\textbf {\bibinfo
  {volume} {106}},\ \bibinfo {pages} {080801} (\bibinfo {year}
  {2011})}\BibitemShut {NoStop}%
\bibitem [{\citenamefont {Haslinger}\ \emph {et~al.}(2018)\citenamefont
  {Haslinger}, \citenamefont {Jaffe}, \citenamefont {Xu}, \citenamefont
  {Schwartz}, \citenamefont {Sonnleitner}, \citenamefont {Ritsch-Marte},
  \citenamefont {Ritsch},\ and\ \citenamefont {M{\"u}ller}}]{BlackbodyPH}%
  \BibitemOpen
  \bibfield  {author} {\bibinfo {author} {\bibfnamefont {P.}~\bibnamefont
  {Haslinger}}, \bibinfo {author} {\bibfnamefont {M.}~\bibnamefont {Jaffe}},
  \bibinfo {author} {\bibfnamefont {V.}~\bibnamefont {Xu}}, \bibinfo {author}
  {\bibfnamefont {O.}~\bibnamefont {Schwartz}}, \bibinfo {author}
  {\bibfnamefont {M.}~\bibnamefont {Sonnleitner}}, \bibinfo {author}
  {\bibfnamefont {M.}~\bibnamefont {Ritsch-Marte}}, \bibinfo {author}
  {\bibfnamefont {H.}~\bibnamefont {Ritsch}}, \ and\ \bibinfo {author}
  {\bibfnamefont {H.}~\bibnamefont {M{\"u}ller}},\ }\href {\doibase
  10.1038/s41567-017-0004-9} {\bibfield  {journal} {\bibinfo  {journal} {Nature
  Physics}\ }\textbf {\bibinfo {volume} {14}},\ \bibinfo {pages} {257}
  (\bibinfo {year} {2018})}\BibitemShut {NoStop}%
\bibitem [{\citenamefont {Nobili}(2016)}]{GammaNobili}%
  \BibitemOpen
  \bibfield  {author} {\bibinfo {author} {\bibfnamefont {A.~M.}\ \bibnamefont
  {Nobili}},\ }\href {\doibase 10.1103/PhysRevA.93.023617} {\bibfield
  {journal} {\bibinfo  {journal} {Phys. Rev. A}\ }\textbf {\bibinfo {volume}
  {93}},\ \bibinfo {pages} {023617} (\bibinfo {year} {2016})}\BibitemShut
  {NoStop}%
\bibitem [{\citenamefont {Williams}\ \emph {et~al.}(2016)\citenamefont
  {Williams}, \citenamefont {wey Chiow}, \citenamefont {Yu},\ and\
  \citenamefont {M\"uller}}]{GammaJason}%
  \BibitemOpen
  \bibfield  {author} {\bibinfo {author} {\bibfnamefont {J.}~\bibnamefont
  {Williams}}, \bibinfo {author} {\bibfnamefont {S.}~\bibnamefont {wey Chiow}},
  \bibinfo {author} {\bibfnamefont {N.}~\bibnamefont {Yu}}, \ and\ \bibinfo
  {author} {\bibfnamefont {H.}~\bibnamefont {M\"uller}},\ }\href
  {http://stacks.iop.org/1367-2630/18/i=2/a=025018} {\bibfield  {journal}
  {\bibinfo  {journal} {New Journal of Physics}\ }\textbf {\bibinfo {volume}
  {18}},\ \bibinfo {pages} {025018} (\bibinfo {year} {2016})}\BibitemShut
  {NoStop}%
\bibitem [{\citenamefont {Chiow}\ \emph {et~al.}(2017)\citenamefont {Chiow},
  \citenamefont {Williams}, \citenamefont {Yu},\ and\ \citenamefont
  {M\"uller}}]{GammaChiow}%
  \BibitemOpen
  \bibfield  {author} {\bibinfo {author} {\bibfnamefont {S.-w.}\ \bibnamefont
  {Chiow}}, \bibinfo {author} {\bibfnamefont {J.}~\bibnamefont {Williams}},
  \bibinfo {author} {\bibfnamefont {N.}~\bibnamefont {Yu}}, \ and\ \bibinfo
  {author} {\bibfnamefont {H.}~\bibnamefont {M\"uller}},\ }\href {\doibase
  10.1103/PhysRevA.95.021603} {\bibfield  {journal} {\bibinfo  {journal} {Phys.
  Rev. A}\ }\textbf {\bibinfo {volume} {95}},\ \bibinfo {pages} {021603}
  (\bibinfo {year} {2017})}\BibitemShut {NoStop}%
\bibitem [{\citenamefont {Roura}(2017)}]{GammaRoura}%
  \BibitemOpen
  \bibfield  {author} {\bibinfo {author} {\bibfnamefont {A.}~\bibnamefont
  {Roura}},\ }\href {\doibase 10.1103/PhysRevLett.118.160401} {\bibfield
  {journal} {\bibinfo  {journal} {Phys. Rev. Lett.}\ }\textbf {\bibinfo
  {volume} {118}},\ \bibinfo {pages} {160401} (\bibinfo {year}
  {2017})}\BibitemShut {NoStop}%
\bibitem [{\citenamefont {Overstreet}\ \emph {et~al.}(2018)\citenamefont
  {Overstreet}, \citenamefont {Asenbaum}, \citenamefont {Kovachy},
  \citenamefont {Notermans}, \citenamefont {Hogan},\ and\ \citenamefont
  {Kasevich}}]{gammaKasevich}%
  \BibitemOpen
  \bibfield  {author} {\bibinfo {author} {\bibfnamefont {C.}~\bibnamefont
  {Overstreet}}, \bibinfo {author} {\bibfnamefont {P.}~\bibnamefont
  {Asenbaum}}, \bibinfo {author} {\bibfnamefont {T.}~\bibnamefont {Kovachy}},
  \bibinfo {author} {\bibfnamefont {R.}~\bibnamefont {Notermans}}, \bibinfo
  {author} {\bibfnamefont {J.~M.}\ \bibnamefont {Hogan}}, \ and\ \bibinfo
  {author} {\bibfnamefont {M.~A.}\ \bibnamefont {Kasevich}},\ }\href {\doibase
  10.1103/PhysRevLett.120.183604} {\bibfield  {journal} {\bibinfo  {journal}
  {Phys. Rev. Lett.}\ }\textbf {\bibinfo {volume} {120}},\ \bibinfo {pages}
  {183604} (\bibinfo {year} {2018})}\BibitemShut {NoStop}%
\bibitem [{\citenamefont {Estey}\ \emph {et~al.}(2015)\citenamefont {Estey},
  \citenamefont {Yu}, \citenamefont {M\"uller}, \citenamefont {Kuan},\ and\
  \citenamefont {Lan}}]{DiffractionBE}%
  \BibitemOpen
  \bibfield  {author} {\bibinfo {author} {\bibfnamefont {B.}~\bibnamefont
  {Estey}}, \bibinfo {author} {\bibfnamefont {C.}~\bibnamefont {Yu}}, \bibinfo
  {author} {\bibfnamefont {H.}~\bibnamefont {M\"uller}}, \bibinfo {author}
  {\bibfnamefont {P.-C.}\ \bibnamefont {Kuan}}, \ and\ \bibinfo {author}
  {\bibfnamefont {S.-Y.}\ \bibnamefont {Lan}},\ }\href {\doibase
  10.1103/PhysRevLett.115.083002} {\bibfield  {journal} {\bibinfo  {journal}
  {Phys. Rev. Lett.}\ }\textbf {\bibinfo {volume} {115}},\ \bibinfo {pages}
  {083002} (\bibinfo {year} {2015})}\BibitemShut {NoStop}%
\bibitem [{\citenamefont {Parker}\ \emph {et~al.}(2016)\citenamefont {Parker},
  \citenamefont {Yu}, \citenamefont {Estey}, \citenamefont {Zhong},
  \citenamefont {Huang},\ and\ \citenamefont {M\"uller}}]{DiffractionRP}%
  \BibitemOpen
  \bibfield  {author} {\bibinfo {author} {\bibfnamefont {R.~H.}\ \bibnamefont
  {Parker}}, \bibinfo {author} {\bibfnamefont {C.}~\bibnamefont {Yu}}, \bibinfo
  {author} {\bibfnamefont {B.}~\bibnamefont {Estey}}, \bibinfo {author}
  {\bibfnamefont {W.}~\bibnamefont {Zhong}}, \bibinfo {author} {\bibfnamefont
  {E.}~\bibnamefont {Huang}}, \ and\ \bibinfo {author} {\bibfnamefont
  {H.}~\bibnamefont {M\"uller}},\ }\href {\doibase 10.1103/PhysRevA.94.053618}
  {\bibfield  {journal} {\bibinfo  {journal} {Phys. Rev. A}\ }\textbf {\bibinfo
  {volume} {94}},\ \bibinfo {pages} {053618} (\bibinfo {year}
  {2016})}\BibitemShut {NoStop}%
\bibitem [{\citenamefont {Peters}\ \emph {et~al.}(2001)\citenamefont {Peters},
  \citenamefont {Chung},\ and\ \citenamefont {Chu}}]{AIChu}%
  \BibitemOpen
  \bibfield  {author} {\bibinfo {author} {\bibfnamefont {A.}~\bibnamefont
  {Peters}}, \bibinfo {author} {\bibfnamefont {K.~Y.}\ \bibnamefont {Chung}}, \
  and\ \bibinfo {author} {\bibfnamefont {S.}~\bibnamefont {Chu}},\ }\href
  {http://stacks.iop.org/0026-1394/38/i=1/a=4} {\bibfield  {journal} {\bibinfo
  {journal} {Metrologia}\ }\textbf {\bibinfo {volume} {38}},\ \bibinfo {pages}
  {25} (\bibinfo {year} {2001})}\BibitemShut {NoStop}%
\bibitem [{\citenamefont {Raab}\ \emph {et~al.}(1987)\citenamefont {Raab},
  \citenamefont {Prentiss}, \citenamefont {Cable}, \citenamefont {Chu},\ and\
  \citenamefont {Pritchard}}]{MolassesPritchard}%
  \BibitemOpen
  \bibfield  {author} {\bibinfo {author} {\bibfnamefont {E.~L.}\ \bibnamefont
  {Raab}}, \bibinfo {author} {\bibfnamefont {M.}~\bibnamefont {Prentiss}},
  \bibinfo {author} {\bibfnamefont {A.}~\bibnamefont {Cable}}, \bibinfo
  {author} {\bibfnamefont {S.}~\bibnamefont {Chu}}, \ and\ \bibinfo {author}
  {\bibfnamefont {D.~E.}\ \bibnamefont {Pritchard}},\ }\href {\doibase
  10.1103/PhysRevLett.59.2631} {\bibfield  {journal} {\bibinfo  {journal}
  {Phys. Rev. Lett.}\ }\textbf {\bibinfo {volume} {59}},\ \bibinfo {pages}
  {2631} (\bibinfo {year} {1987})}\BibitemShut {NoStop}%
\bibitem [{\citenamefont {Vuleti\ifmmode~\acute{c}\else \'{c}\fi{}}\ \emph
  {et~al.}(1998)\citenamefont {Vuleti\ifmmode~\acute{c}\else \'{c}\fi{}},
  \citenamefont {Chin}, \citenamefont {Kerman},\ and\ \citenamefont
  {Chu}}]{RSCChu}%
  \BibitemOpen
  \bibfield  {author} {\bibinfo {author} {\bibfnamefont {V.}~\bibnamefont
  {Vuleti\ifmmode~\acute{c}\else \'{c}\fi{}}}, \bibinfo {author} {\bibfnamefont
  {C.}~\bibnamefont {Chin}}, \bibinfo {author} {\bibfnamefont {A.~J.}\
  \bibnamefont {Kerman}}, \ and\ \bibinfo {author} {\bibfnamefont
  {S.}~\bibnamefont {Chu}},\ }\href {\doibase 10.1103/PhysRevLett.81.5768}
  {\bibfield  {journal} {\bibinfo  {journal} {Phys. Rev. Lett.}\ }\textbf
  {\bibinfo {volume} {81}},\ \bibinfo {pages} {5768} (\bibinfo {year}
  {1998})}\BibitemShut {NoStop}%
\bibitem [{\citenamefont {Fitzgibbon}\ \emph {et~al.}(1999)\citenamefont
  {Fitzgibbon}, \citenamefont {Pilu},\ and\ \citenamefont
  {Fisher}}]{EllipseFisher}%
  \BibitemOpen
  \bibfield  {author} {\bibinfo {author} {\bibfnamefont {A.}~\bibnamefont
  {Fitzgibbon}}, \bibinfo {author} {\bibfnamefont {M.}~\bibnamefont {Pilu}}, \
  and\ \bibinfo {author} {\bibfnamefont {R.~B.}\ \bibnamefont {Fisher}},\
  }\href {\doibase 10.1109/34.765658} {\bibfield  {journal} {\bibinfo
  {journal} {IEEE Transactions on Pattern Analysis and Machine Intelligence}\
  }\textbf {\bibinfo {volume} {21}},\ \bibinfo {pages} {476} (\bibinfo {year}
  {1999})}\BibitemShut {NoStop}%
\bibitem [{\citenamefont {Foster}\ \emph {et~al.}(2002)\citenamefont {Foster},
  \citenamefont {Fixler}, \citenamefont {McGuirk},\ and\ \citenamefont
  {Kasevich}}]{EllipseKasevich}%
  \BibitemOpen
  \bibfield  {author} {\bibinfo {author} {\bibfnamefont {G.~T.}\ \bibnamefont
  {Foster}}, \bibinfo {author} {\bibfnamefont {J.~B.}\ \bibnamefont {Fixler}},
  \bibinfo {author} {\bibfnamefont {J.~M.}\ \bibnamefont {McGuirk}}, \ and\
  \bibinfo {author} {\bibfnamefont {M.~A.}\ \bibnamefont {Kasevich}},\ }\href
  {\doibase 10.1364/OL.27.000951} {\bibfield  {journal} {\bibinfo  {journal}
  {Opt. Lett.}\ }\textbf {\bibinfo {volume} {27}},\ \bibinfo {pages} {951}
  (\bibinfo {year} {2002})}\BibitemShut {NoStop}%
\bibitem [{\citenamefont {Yu~et. al.}()}]{Chenghui}%
  \BibitemOpen
  \bibfield  {author} {\bibinfo {author} {\bibfnamefont {C.}~\bibnamefont
  {Yu~et. al.}},\ }\href@noop {} {\bibinfo  {journal} {to be published}\
  }\BibitemShut {NoStop}%
\end{thebibliography}
\end{document}